\documentclass{aa}  
\usepackage[dvipsnames]{xcolor}

\usepackage{graphicx}
\usepackage{txfonts}
\usepackage{xcolor}
\usepackage{mathrsfs}
\usepackage[colorlinks=true, citecolor=blue]{hyperref}
\usepackage{physics}
\usepackage{nicefrac}
\usepackage[normalem]{ulem}
\usepackage{soul}
\usepackage[colorlinks=true, citecolor=blue]{hyperref}
\usepackage{amsmath}
\usepackage{lastpage}
\usepackage{derivative}

\usepackage[version=4,arrows=pgf-filled,
textfontname=sffamily,
mathfontname=mathsf]{mhchem}

\newcommand{\Laplace}{\nabla^2}
\newcommand{\mn}[1]{\langle#1\rangle}

\newcommand{\NIRVANA}{\textsc{nirvana-iii}\xspace}
\newcommand{\PDFFT}{\textsc{p}{\tiny{3}}\textsc{dfft}\xspace}

\begin{document} 

\title{An efficient spectral Poisson solver for the \NIRVANA code: the shearing-box case with vertical vacuum boundary conditions}

\author{S. Rendon Restrepo \inst{1}\thanks{\email{srendon@aip.de}}
        \and O. Gressel \inst{1,2}}

\institute{Leibniz-Institut für Astrophysik Potsdam (AIP), An der Sternwarte 16, D-14482 Potsdam, Germany
\and Niels Bohr International Academy, Niels Bohr Institute, Blegdamsvej 17, DK-2100 Copenhagen Ø, Denmark 
}

\titlerunning{A free-space spectral Poisson solver for the \NIRVANA code}

\authorrunning{Rendon Restrepo \& Gressel}

\abstract
{The stability of a differentially rotating fluid subject to its own gravity is a problem with applications across wide areas of astrophysics -- from protoplanetary discs to entire galaxies. The shearing box formalism offers a conceptually simple framework for studying differential rotation in the local approximation.
}
{Aimed at self-gravitating, and importantly, vertically stratified protoplanetary discs, we develop two novel methods for solving Poisson's equation in the framework of the shearing box with vertical vacuum boundary conditions (BCs).
}
{
Both approaches naturally make use of multi-dimensional fast Fourier transforms for computational efficiency. While the first one exploits the linearity properties of the Poisson equation, the second, which is slightly more accurate, consists of finding the adequate discrete Green's function (in Fourier space) adapted to the problem at hand. 
To this end, we have derived, in Fourier space, an analytical Green's function satisfying the shear-periodic BCs in the plane as well as vacuum BCs, vertically.
}
{Our spectral method demonstrates excellent accuracy, even with a modest number of grid points, and exhibits third-order convergence.
It has been implemented in the \NIRVANA code, where it exhibits good scalability up to 4096 CPU cores, consuming less than 6\% of the total runtime. 
This was achieved through the use of \PDFFT, a fast Fourier Transform library that employs pencil decomposition, overcoming the scalability limitations inherent in libraries using slab decomposition.
}
{We have introduced two novel spectral Poisson solvers that guarantees high accuracy, performance, and intrinsically support vertical vacuum boundary conditions in the shearing-box framework. Our solvers enable high-resolution local studies involving self-gravity, such as MHD simulations of gravito-turbulence and/or gravitational fragmentation.
}

\keywords{self-gravity --
          numerical methods --
          spectral methods -- Poisson equation}

\maketitle
%

\section{Introduction}

Many astrophysical phenomena require accounting for the gravity of a dilute gas or plasma, often referred to as self-gravity (SG). 
This includes processes such as the core collapse of molecular clouds \citep{1977_shu, 1980_norman, 2023_he_ricotti}, episodic outbursts in FU Orionis stars \citep{2001_armitage, 2006_vorobyov}, and accretion and angular momentum transport in young protoplanetary discs, which can lead to fragmentation for efficient cooling \citep{2001_gammie, 2016_Kratter_Lodato, 2021b_baehr_zhu}.
Recent theoretical and numerical studies in the local frame, also known as the shearing box approximation \citep[see, e.g.,][]{2007_gressel,2010_stone}, have demonstrated the potential of gravitational instability to amplify weak magnetic fields through a process called the gravitoturbulent dynamo \citep{2019_riols,2020_deng, 2023_lohnert}.

The numerical computation of SG is crucial for these processes, necessitating to solve the Poisson equation,
\begin{equation}\label{Eq: poisson}
\Laplace \Phi = 4 \pi G \rho,
\end{equation}
where $\Phi$ and $\rho$ represent the gravitational potential and volume density, respectively. 
A standard approach in Cartesian uniform grids involves discretizing this equation, leading to a linear system typically solved using iterative multigrid methods, as demonstrated by \citet{2005_ziegler} in the finite-volume MHD code \NIRVANA. 
Nonetheless, this requires evaluating accurately the SG potential at the numerical domain boundaries, which generally demands to resort to a multipole expansion.
A more accurate method is the screening method \citep{1977_james,2019_moon,2024_gressel}, which enables computing the exact potential at the domain boundaries.

The numerical computation of the gravitational potential can also be performed using spectral methods.
These techniques often involve solving the Poisson equation in Fourier space,
\begin{equation}\label{Eq: poisson spectral}
\hat{\Phi}(\vec{k}) = 
\displaystyle - \frac{4\pi G\, \hat{\rho}}{||\vec{k}||^2}  \quad (\mathrm{for}\ \vec{k}\ne\vec{0}) 
\,,
\end{equation}
followed by an inverse transform in order to obtain the potential in real space. 
Their primary advantages are accuracy and efficiency, and --with an optimum numerical complexity of $N \log \left(N\right)$-- they are well-suited for large-scale problems. 
The spectral approach, however, relies on the discrete Fourier transform (DFT), or, more precisely, the fast Fourier transform (FFT) -- where the crucial advantage of the latter is that it overcomes the quadratic complexity order of naive DFT algorithms in favour of the above mentioned scaling. A central assumption in such discrete methods, however, is that the input signal represents one complete period of a periodic signal. 
This is a significant limitation for gravitational problems, as it implies the existence of infinite copies of the numerical window in all directions, which is often unrealistic or impractical. 
Additionally, it requires, for consistency, that the volume integral of the source term of the Poisson equation vanishes \citep{2008_binney_tremaine, 2023_mandal}. 
Therefore, it becomes desirable to develop, in Cartesian grids, a full spectral solver that not only ensures spectral accuracy and efficiency, but also captures unbound -- or ``free-space'' -- boundary conditions for the potential.

The Hockney-Eastwood method, which is second-order accurate \citep{1981_hockney}, meets most of the above constraints, except for spectral accuracy.
Another method, proposed by \citet{2016_vico_greengard_ferrando} (thereafter VGF), and which is widely known in plasma physics and condensed matter, involves modifying the Green's function to account for the unbound nature of the potential outside the numerical domain. This method, however, appears to have been overlooked by the astrophysics community.
Most importantly, under the VGF formalism the Green's function in Fourier space has an analytical form and is regularized at the singularity, making it suitable for FFT methods. 
Benchmarks have shown that the relative errors of this method reach machine accuracy (i.e., for uniform grids) already at a modest number of points, outperforming any fixed order of convergence algorithm \citep{2021_zou, 2024_mayani}.
However, the VGF method is not yet adapted to the shearing box approximation \citep{1965_goldreich}, that is, for the case when the vertical structure has to be included \citep[but see][for a complementary approach]{2009_koyama}. 
In our case, the box features two periodic dimensions (i.e., in the horizontal plane) as well as one free-space boundary condition (i.e., in the vertical dimension).

In this short article, we present two new spectral methods for the Poisson equation intended to be used in the Cartesian geometry. 
We demonstrate convergence and performance for our reference implementation within the finite-volume code \NIRVANA.
Specifically, we detail the derivation and implementation of a new analytical Green's function meant for 3D shearing box simulations.
We begin describing our new spectral solver, and its numerical specifics in Sect.~\ref{sec: general description}.
In Sect.~\ref{sec: accuracy tests}, we benchmark the accuracy of our solver with two static and one dynamic test.
Finally, in Sect.~\ref{sec: performance} we quantify the performance of our solver employing distributed-memory parallelism with standard message passing interface (MPI) routines.

\section{General description}\label{sec: general description}

In the following, we will discuss some basic considerations that will provide the foundation for our spectral Poisson solver.
We then introduce two novel methods--which, to our knowledge, are new for studying SG in stratified discs.

\subsection{Mapping to fully periodic points}

The methods discussed in this paper require periodicity in both the $x$- and $y$-directions.
In the shearing-box, where the $x$ boundaries are permanently in motion, strict periodicity is only satisfied in the $y$-direction, and during special points in time for the $x$-direction.
To address this, one first needs to perform a coordinate transformation along the $y$-direction \citep{2001_gammie}, either using a simple linear interpolation \citep{2007_gressel} or the so-called ``shear advection by Fourier interpolation'' (SAFI) scheme \citep{2009_johansen}.
It is important to note that this mapping introduces a coordinate transformation that must be accounted for in the Poisson equation \citep{2023_zier}, modifying the wave-vector in the periodic frame as follows:
\begin{equation}
\vec{k}(t)=
\left(
\begin{array}{l}
k_x \\
k_y \\
k_z
\end{array}
\right)
=
\left(
\begin{array}{c}
K_x + \frac{\Delta y_0 (t)}{L_x} K_y \\
K_y \\
K_z
\end{array}
\right)
\end{equation}
where $\Delta y_0(t)=\text{mod}\left(\frac{3}{2} \Omega L_x t, L_y \right)$ and $\vec{K}$ is the wave-vector in the initial frame.
Please note, that until the end of the paper, we will refer only to the periodic frame and its associated wave-vector $\vec{k}$.
Since $\Delta y_0/L_x$ periodically varies between 0 and $L_y/L_x$, the wave-vector space is a rectangular cuboid whose x length varies continuously between $K_x$ and $K_x + \frac{L_y}{L_x} K_y$.
Second, the Poisson equation is then solved in this new, fully-periodic (i.e., in the horizontal plane) reference frame. 
Finally, the obtained potential is subsequently mapped back to the original coordinate system, via the reverse mapping. 

In the following sections, we will primarily focus on the second step of the previously outlined scheme: solving the Poisson equation with two periodic boundary conditions (BCs) and a vacuum boundary condition in the vertical direction. Henceforth, any reference to wave-vectors will pertain to $\vec{k}$.
We now present our two techniques based on the spectral approach to solving Poisson's equation.

\subsection{The Superposition Analytical-Spectral Hybrid Approach (SASHA)}
\label{subsec: superposition technique}

In this section, we aim to solve equation Eq.~(\ref{Eq: poisson}), subject to mixed BCs. 
By invoking the principle of superposition, we can decompose Poisson's equation into a constant term and a deviation from this zeroth-order term,
\begin{equation}
\Laplace \Phi = 4 \pi G \, \mn{\rho} + 4 \pi G\, \left(\rho - \mn{\rho}\right),
\end{equation}
where $\mn{\rho}$ is the volume-averaged mass density.
First, we solve $\Laplace \Phi_0 = 4 \pi G \mn{\rho}$, which can be integrated subject to the boundary conditions and symmetry requirements. As can be checked, a straightforward solution is
\begin{equation}
\Phi_0(z) = 4 \pi G \mn{\rho}\, z\, \left( \frac{1}{2} z - z_0 \right)\,,
\end{equation}
where $z_0=\mn{z\,\rho}/\mn{\rho}$ is the vertical coordinate of the centre of mass.
In the second step, we utilize the numerically efficient spectral method, as presented in Eqn.~(\ref{Eq: poisson spectral}), to solve for $\Laplace \Phi_{\text{P}} = 4 \pi G \, \left(\rho - \mn{\rho}\right)$. 
Specifically, DFTs require periodic boundary conditions in all directions, which is only satisfied only in $x$ and $y$ directions.
Therefore, we used an aperiodic convolution in the vertical direction, which simply requires to artificially make the domain periodic in the vertical direction by the use of the zero-padding technique \citep{1981_hockney}.
This technique consists of doubling the input signal with zeros, allowing periodic boxes in the vertical direction to not affect the computational domain and improving the frequency resolution in Fourier space. 
As we will see in Section \ref{sec: accuracy tests}, this does not alter the solution.
We stress that the requirement of periodic boundary conditions in all directions implies that the volume integral of the left-hand side of the Poisson equation must vanish:
\begin{equation}\label{Eq: poisson consistency}
\iiint \vec{\nabla} \cdot \vec{\nabla} \Phi_P \, d^3 \vec{r}  = \iint \frac{\partial \Phi_P}{\partial \vec{n}} d^2S = 0.
\end{equation}
Therefore, the subtraction of the mean density from the source term for this subproblem ensures mathematical consistency \citep{2008_binney_tremaine, 2023_mandal}.
In the final step, the full gravitational potential in the shearing box is obtained as the sum of the two aforementioned potentials, $\Phi = \Phi_0 + \Phi_P$, and is consistent with the treatment of gravity.

\subsection{The Vico-Greengard-Ferrando with Hybrid Boundary Conditions (VGF-HybridBC) method}\label{subsec: VGF revisited}

Before delving into our second method, it is instructive to study the one-dimensional case first.
Here, the mass density, $\rho(z)$, is assumed to be constant along each (infinte) horizontal slab, and hence only depends on the vertical coordinate, $z$. 
 In this case, the Laplacian becomes $\Laplace= \dv[2]{}{z}$, and the PDE is transformed into a simple ODE,
\begin{equation}
\dv[2]{\Phi}{z} =  4\pi G\, \rho(z)\,,
\end{equation}
which can be integrated using a convolution approach.
This procedure employs the inhomogeneous 1D Green's function
\begin{equation}\label{eq: one-dim-green}
\mathcal{G}_{1D}(z,z') =  \frac{1}{2} |z-z'|\, ,
\end{equation}
which produces the solution $\Phi(z)=4\pi\,G\; \int \mathcal{G}_{1D}(z,z')\,\rho(z')\,{\rm d}z'$.
Contrary to methods using the homogeneous Green's function, which necessitate the potential to be augmented by a linear function in $z$, this procedure yields the correct symmetry and asymptotic behaviour.
Indeed, the force norm must become symmetric when extending far above and below the midplane\footnote{For the simple case of a constant density, it can be shown that this is equivalent of a parabola with its vertex at the centre of mass location (also cf. the argument in Sect.~\ref{subsec: superposition technique}, above).}.
Specifically, in the asymptotic limit (i.e., at great height above\,/\,below the midplane) the potential, $\Phi(z)$, becomes a linear function -- resulting in a constant gravitational acceleration. 
This is in contrast to the 3D case, where the acceleration drops with distance. 
The explanation is in the assumed infinite extend of each slab of material -- the further one departs from the midplane, the larger part of the density distribution can be ``seen'' (under a constant opening angle) at any given location. Since the seen area grows quadratic with height, the quadratic dependence on separation --the very hallmark of Newtonian gravity-- gets entirely compensated in that case.

Extending this approach, we now introduce a Fourier-based convolution method that is inherently three-dimensional and fully compatible with vertical vacuum boundary conditions.
We aim to derive the appropriate Green's function for the Poisson equation, adapting the VGF method by \citet{2016_vico_greengard_ferrando} to a scenario with two periodic boundary conditions and one vacuum boundary condition.
Given the periodic nature of the problem in the $x$ and $y$ directions, we apply a Fourier transform in these dimensions, resulting in a 1D Helmholtz equation (or screened Poisson equation) in the $z$-direction:
\begin{equation}\label{Eq: helmholtz equation}
\left(\dv[2]{z} - k^2 \right) \Tilde{\Phi}(z) = 4 \pi G \tilde{\rho}(z)
\end{equation}
where $k^2 = k_x^2+k_y^2$, and $\tilde{\Phi}$ and $\tilde{\rho}$ represent the partial Fourier transforms of the potential and density in the $x$ and $y$ directions, respectively.
The associated Green's function to the above linear differential equation is:

\begin{equation}\label{Eq: green's function original}
 \mathcal{G}_k(z-z') = 
 \left\{
\begin{array}{cll}
\displaystyle\frac{1}{2} |z-z'|                & \text{if} & k=0        \\ [8pt]
\displaystyle-\frac{1}{2 k} e^{-k\,|z-z'|}     & \text{if} & k \neq 0
\end{array}
 \right.
\end{equation}

We observe a discontinuity in the Green's function as $k \rightarrow 0$.
Indeed, for $k=0$, the Green's function diverges linearly at infinity, indicating a net monopole (uncanceled zero-mode) and reflecting a constant gravitational field (see previous paragraph).
Conversely, the $k \neq 0$ Green's function mitigates this divergence by introducing screening.
Both cases combined permit to satisfy physical boundary conditions at infinity (in the vertical direction) for a stratified disc. 
This Green's function enables us to obtain the potential via convolution: 

\begin{equation}\label{Eq: convolution original}
\begin{array}{ll}
\tilde{\Phi}(z) & \displaystyle = 4 \pi G \, \int\limits_{-\infty}^{\infty} \mathcal{G}_k(z-z')\; \tilde{\rho}(z') \,{\rm d}z'\,.
\end{array}
\end{equation}

The first trick of the VGF method involves recognizing that the integral form for the potential remains unchanged in a finite-sized box if the Green's function is replaced by:

\begin{equation}\label{Eq: green's function with rect}
\mathcal{G}^L(z-z') = \text{rect}\left(\frac{z-z'}{2 L} \right)\;\mathcal{G}_k(z-z')\,,
\end{equation}

where $L=\alpha L_z$ is a suitable enclosure (choosing $\alpha>1$), $L_z$ is the vertical extent of the actual numerical box, and `rect' is the rectangular function.
For the rest of the paper we adopt a value $\alpha=1.1$ \citep{2024_mayani}.

The second trick involves computing the Fourier transform of the above Green's function in the vertical direction

\begin{equation}\label{Eq: VGF shearing box}
  \hat{\mathcal{G}}^L(k, k_z)  = \int\limits_{-\infty}^{\infty}  \mathcal{G}^L(z)\; e^{i k_z z} \, {\rm d}z\,,
\end{equation}

where three cases must be distinguished (see Appendix \ref{app: Green's function in Fourier space}):

\begin{equation}\label{Eq: Green's function regularized in Fourier}
   = \left\{ \begin{array}{ll}
      \displaystyle {L^2}/{2}  & \text{if} \, k=0\,, k_z=0\,, \\[6pt]
      \displaystyle \frac{\left[ k_z L \sin{(k_z L)} + \cos{(k_z L)} -1 \right]}{k_z^2}   & \text{if} \, k=0\,, k_z\ne 0\,, \\[6pt]
      \displaystyle -\frac{ e^{-k L} \left( \frac{k_z}{k} \sin{(k_z L)} - \cos{(k_z L)} \right) + 1 }{k^2+k_z^2} & \text{otherwise}
      \,.
   \end{array} \right.
\end{equation}
Finally, the potential in real space is simply obtained by an inverse Fourier transform of the product between the Fourier transformed Green's function and density:
\begin{equation}
\Phi(x,y,z) = 4 \pi G \, \mathcal{F}^{-1} \left( \hat{G}^L \cdot \hat{\rho} \right)
\end{equation}
where $\hat{\rho}$ is the full Fourier transform of the density.
It is noteworthy that our calculation naturally resolves the Green's function singularity for $(k, k_z)=(0,0)$.

\subsection{Implementation details}

It is customary to use the FFTW3 library \citep{FFTW05} for spectral methods; 
however, it only allows parallelization in one direction, affecting globally the performance of the hydrodynamical solver of \NIRVANA.
Indeed, the granularity of the hydrodynamical solver is optimal for a 3D parallel decomposition.
For this reason, we opted for the \PDFFT library \citep{2012_p3dfft}, which uses a "pencil" decomposition (i.e., along two out of three spatial dimensions, leaving the third local in memory). This is a good compromise, balancing the performance of the hydro code and the spectral gravitational solver.
The \PDFFT version uses stride-1 decomposition, transforming $(Z,X,Y)$ arrays directly into $(Y,X,Z)$ complex arrays—eliminating MPI\_ALL\_TO\_ALL calls and transpositions, resulting in significant time and memory savings.
In this context, we would like to mention that \citet{2009_koyama} presented an FFT method applicable to shearing boxes.
However, it requires to proceed in four steps: 1)~in-plane Fourier transform, 2)~modification of the density in the mirror regions and vertical FFT, 3)~convolution, and 4)~backward transform.
This particular procedure, specifically the decomposition of the 3D transform in two steps precludes the use of parallel three dimensional FFT, hindering the use of \PDFFT and therefore limiting the performance.
The advantage of our method lies that the Fourier transform in the three directions is done in a single step.

Because of the vacuum BCs in the vertical direction, the SASHA method for linear decomposition, discussed in Section~\ref{subsec: superposition technique}, employs a standard ``zero-padding'' (ZP) technique. 
Although ZP (applied along one dimension only) doubles the memory usage, the Hermitian property of the Fourier transforms of real quantities reduces the number of complex elements by half, compensating for this effect. 
Additionally, most FFT algorithms are optimized for sizes that are multiples of 2.
The VGF-HybridBC method highlighted in Section \ref{subsec: VGF revisited} also necessitates padding, but it requires quadrupling the domain in the vertical direction.
This is due to handling aperiodic convolution in the vertical direction and the oscillatory nature of $\mathcal{G}^L$ \citep{2016_vico_greengard_ferrando, 2024_mayani}. 
To address this issue, \citet{2016_vico_greengard_ferrando} proposed a method to reduce the grid size from quadruple to double by using a pre-computation step. 
However, this approach is not feasible in our specific case due to the time-dependent nature of the Green's function, making any pre-computation impractical. 
We note that ZP can in principle be combined with so-called ``pruned transforms'', which makes the FFT library aware of the regions filled with zeros, resulting in memory savings and avoiding certain unnecessary communications between processors.
However, in \PDFFT, this filtering is only possible for the wave-spectrum data and not the real data. 
Indeed, we speculate that the ``pruned transform'' module of \PDFFT was likely created with hydrodynamical turbulence applications in mind, where the 2/3 filtering rule is used to avoid aliasing.

For production runs, a crucial consideration is determining the potential value in ghost cells in order to compute forces.
Initially, it was tempting to use the gravitational potential from the zero-padded region as the buffer zone.
However, these values lack of physical meaning and cannot be related to actual data.
Therefore, it was decided to ensure that the gradient of the full potential, $\Phi$, in the ghost cells matches that of the last active cell in vertical direction.

\section{Tests and numerical convergence}\label{sec: accuracy tests}

In this section we benchmark the accuracy of our method and check our numerical implementation with two static and one dynamic test, respectively.
We start by retrieving the potential associated with a Gaussian vertical distribution.

\subsection{1D vertical test}\label{subsec: 1D vertical test}

\begin{figure}
\centering
\includegraphics[width=\hsize]{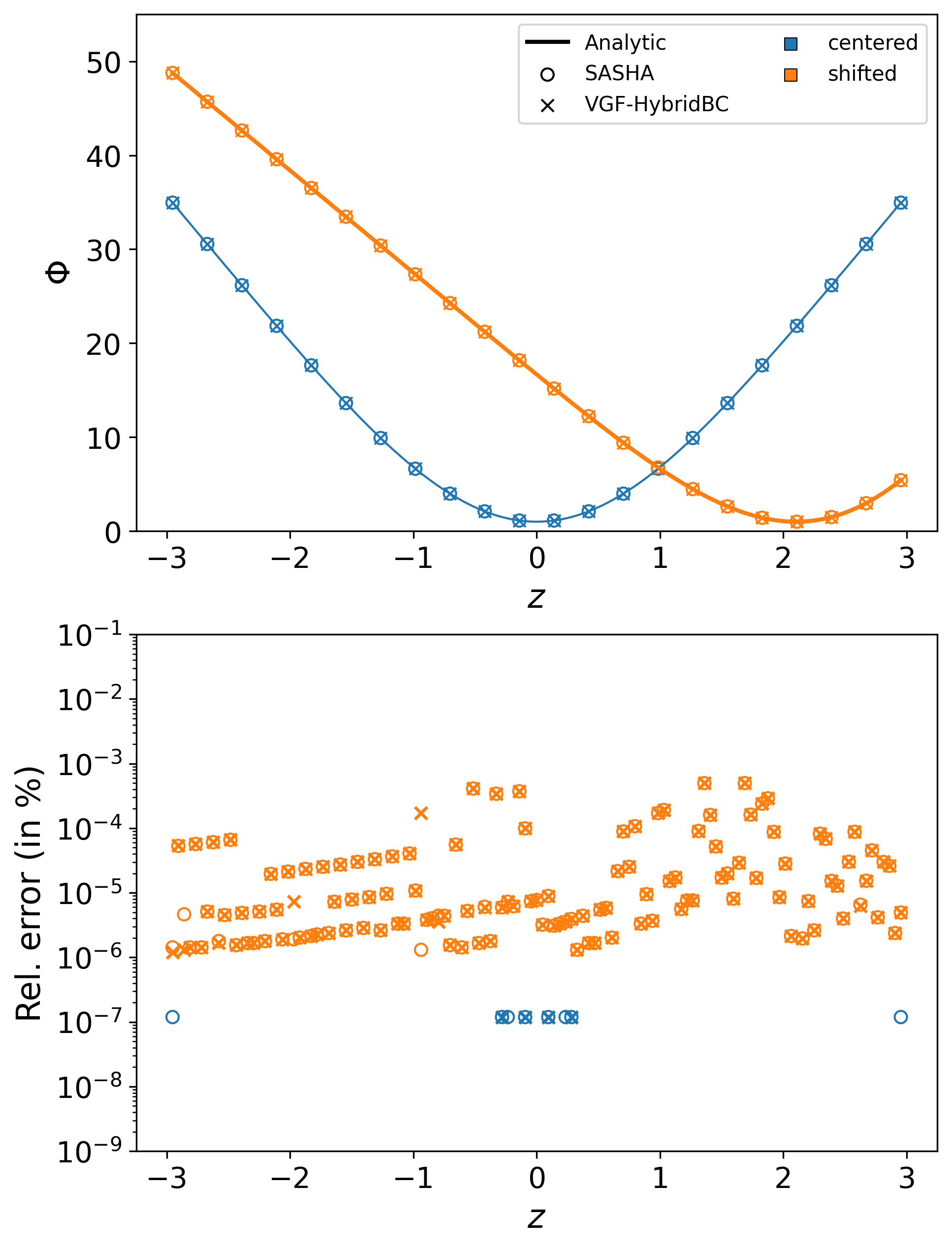}

\caption{Potential associated with a one-dimensional vertical Gaussian distribution.
We considered two cases: ~i) when centred around the midplane, and ~ii) with a shifted Gaussian profile.
}
\label{fig: potential 1d}
\end{figure}

Through this test we aim to to demonstrate that both our methods, coupled with the ZP technique, provide correct and accurate results in a setup where vertical vacuum boundary conditions are considered.
Therefore, we initiate the density profile as a shifted Gaussian,
\begin{equation}
\rho(z) = \left\{
\begin{array}{lll}
 \displaystyle e^{-\frac{1}{2} \left(z-z_0\right)^2}     & \text{if} & z \in [-L_z/2,L_z/2]  \\
 0    &   \text{else} & 
\end{array}
\right.
\end{equation}
where the extent of the numerical window in the vertical direction is $[-L_z/2, L_z/2]$.
Here it is particularly important to also solve Poisson equation in the vacuum regions by ensuring continuity of the potential and it's derivative at the box vertical boundaries and ensuring that the force associated to the potential should be symmetric with respect to $z$ far away from the midplane, that is, $\;{\rm d}\Phi/{\rm d}z\,(-\infty)=-{\rm d}\Phi/{\rm d}z\,(+\infty)$.
With these considerations in mind, the gravitational potential in the numerical window (i.e., $[-L_z/2, L_z/2]$) is:
\begin{equation}
\displaystyle \frac{\Phi(z)}{4 \pi G} = \sqrt{\frac{\pi}{2}} \left(z-z_0\right) \mathrm{erf} \left( \frac{z-z_0}{\sqrt{2}} \right)   + e^{-\frac{1}{2} \left(z-z_0\right)^2} + a\, z\,,
\end{equation}
with 
\begin{equation}
a=-\frac{1}{2} \sqrt{\frac{\pi}{2}} \left[ \text{erf}\left( \frac{-L_z/2-z_0}{\sqrt{2}} \right) + \text{erf}\left( \frac{L_z/2-z_0}{\sqrt{2}} \right) \right]\,,     \\
\end{equation}
and where we took the integration constant equal to zero.

For this test, we choose fiducial parameters $L_x=L_y=L_z=6$, as well as $N_x=N_y=N_z=128$, and $z_0=[0, 2.0]$.
The choice of a uniform cubic grid, although not necessary, allows to infer the limit of optimal accuracy. Similarly, the offset parameter, $z_0$, allows to compare a symmetric, as well as an off-centre density distribution, respectively.

In the top panel of Figure~\ref{fig: potential 1d}, we show the analytic solution for the gravitational potential and compare it with the numerical results provided by both our spectral methods.
For the sake of comparison, in these graphics, we subtracted a constant such that all potentials have their minimum equal to 1.
The analytical prediction and numerical estimates for the centred and shifted distributions are indistinguishable.
More specifically, in the bottom panel of Fig.~\ref{fig: potential 1d} we depicted the relative error,
\begin{equation}
\epsilon = \left| \frac{\Phi_{num}-\Phi_a}{\Phi_a} \right|,
\end{equation}
which ranges between $~10^{-6}$ and $~10^{-4}$ for the shifted distribution.
For the centred distribution, the numerical and analytical potential are identical except in 4 points close to the midplane where the relative error is only about $10^{-7}$.

\subsection{3D test with mixed boundary conditions}\label{subsec: 3d with mixed BC}

\begin{figure}
\centering
\includegraphics[width=0.9\hsize]{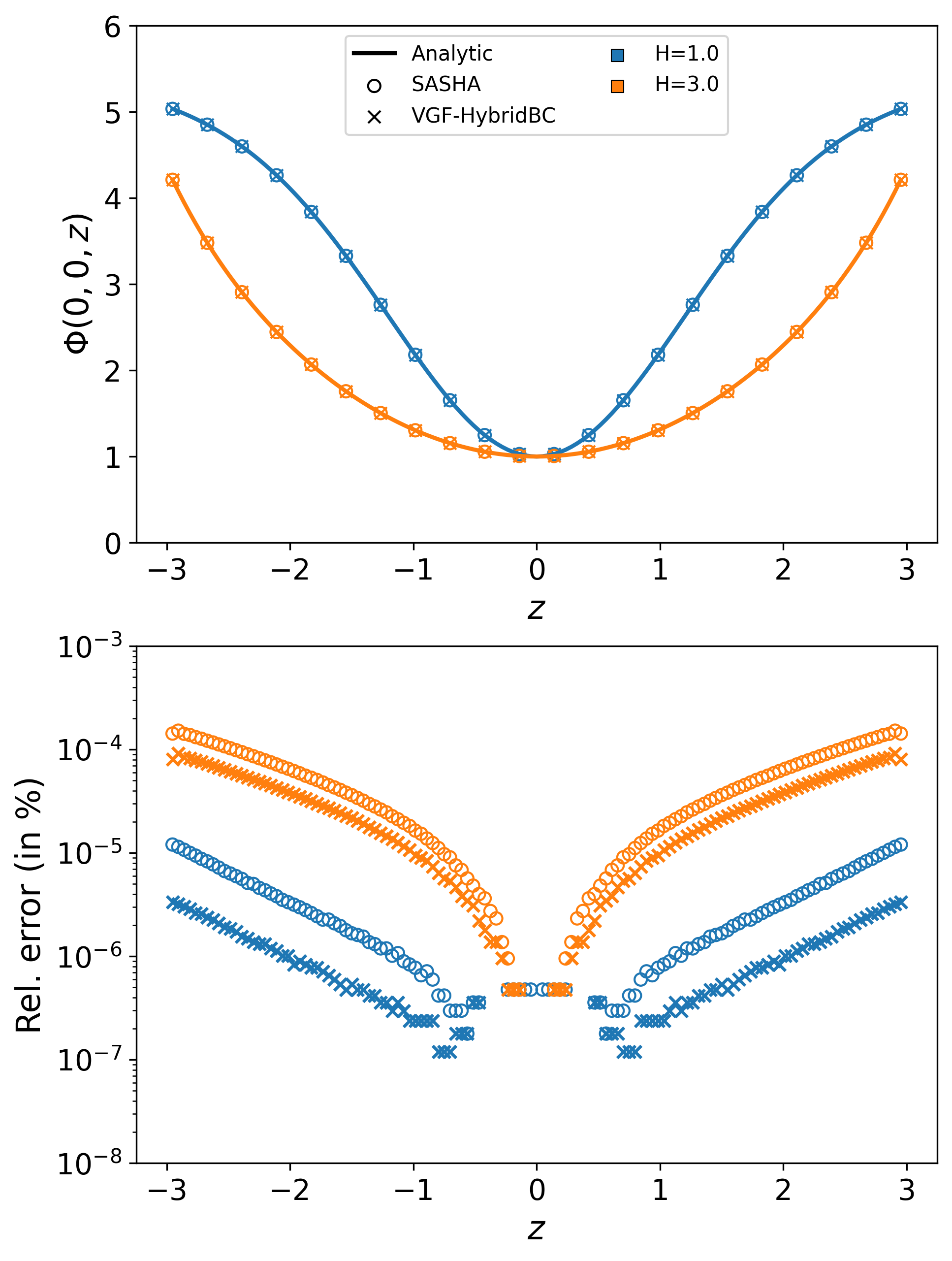}
\caption{Potential along the z direction associated with a 3D density distribution: periodic in x and y, Gaussian in the vertical direction. 
}
\label{fig: potential 3d}
\end{figure}

\begin{figure}
\centering
\includegraphics[width=\hsize]{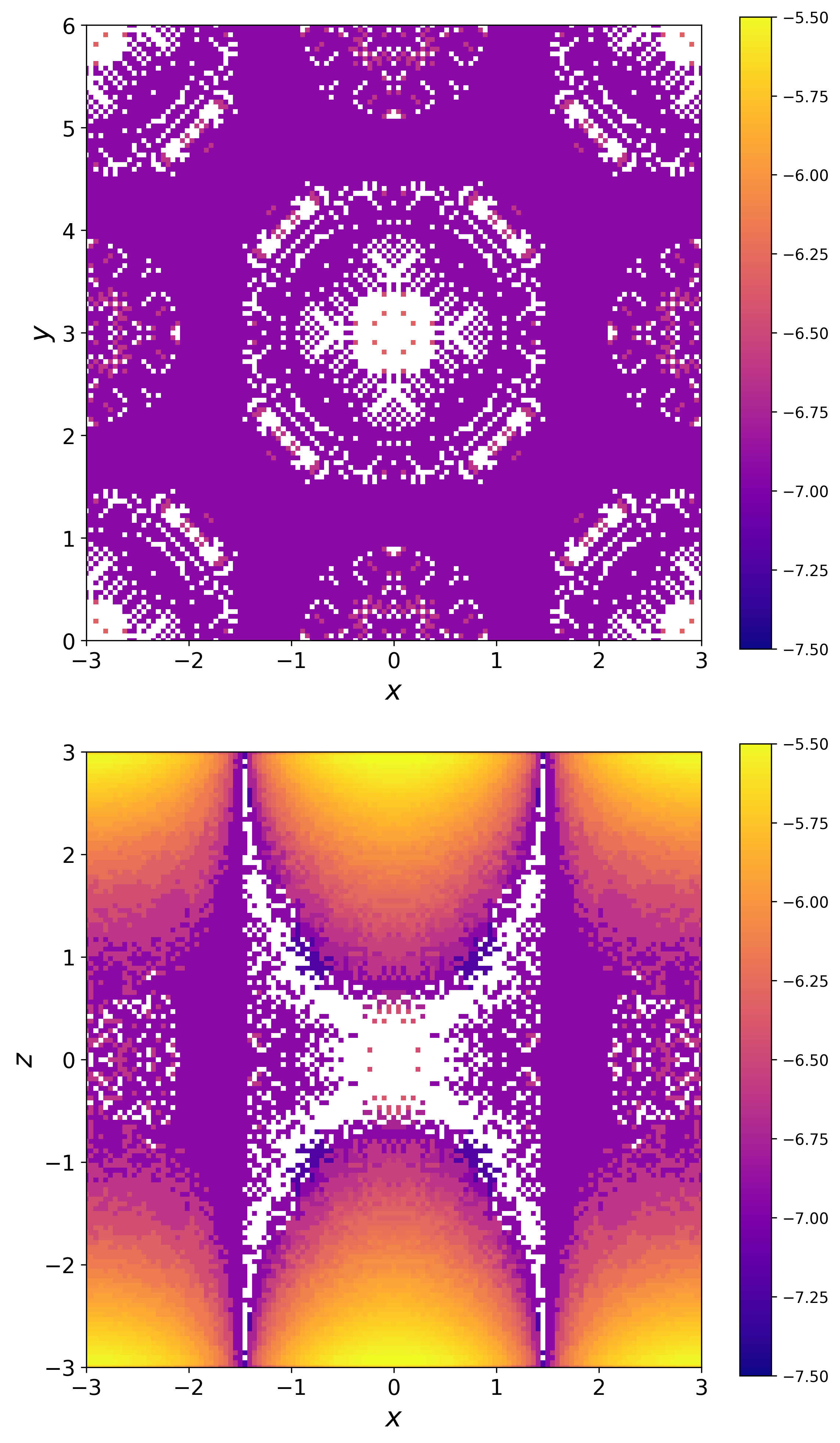}
\caption{Log-scale relative error cuts for the 3D test, employing the VGF-HybridBC method, with mixed boundary conditions.
Top: $z=0$ plane. Bottom: $y=3$ plane.
}
\label{fig: err_xy and err_xz}
\end{figure}

A more challenging test consists on benchmarking our solver in a setup with two periodic boundary conditions and one unbound direction.
Therefore, we chose a density distribution (defined on $z \in [-L_z/2,L_z/2]$),
\begin{equation}
  \rho(x,y,z) = \cos(\vphantom{k_y}k_x x)\, \cos(k_y y)\,  e^{-\frac{1}{2} \left({z}/{H}\right)^2} \,,
\end{equation}
whose associated potential (applicable in the same domain) is:
\begin{equation}
\Phi(x,y,z) = 4\pi G\, \cos(\vphantom{k_y}k_x x)\, \cos(k_y y)\; 
            \Big[  2 a \cosh{(k z)} + F_+ + F_- \Big]\,,  
\end{equation}
with $k = \sqrt{k_x^2 + k_y^2}$, as well as functions
\begin{equation}
\begin{array}{lcl}
F_\pm(z) & = & \sqrt{\frac{\pi}{2}} \frac{H}{2 k}\, e^{\frac{k H}{2} \left( kH \pm 2z/H \right)}\, \text{erf}\left( \frac{kH \pm z/H}{\sqrt{2}}  \right)  \\ [8pt]
a & = & -\sqrt{\frac{\pi}{2}} \frac{H}{2k} e^{\frac{\left( k H \right)^2}{2}} \text{erf} \left( \frac{kH + L_z/(2H)}{\sqrt{2}}  \right)
\end{array}
\end{equation}
As in Sect.~\ref{subsec: 1D vertical test}, we also obtained this solution by solving Poisson’s equation in the vacuum regions and enforcing continuity of both the potential and its derivative at the domain boundaries. 
However, unlike in Section~\ref{subsec: 1D vertical test}, we additionally imposed the condition that the potential vanishes at large distances from the midplane (see Sect. \ref{subsec: VGF revisited}).

For this study, we set $N_x=N_y=N_z=128$, as well as $L_x=L_y=L_z=6$, and choose the scale height a)~three times smaller than the box vertical extent, and b)~ as large as the latter -- in other words $H \in [1.0, 3.0]$. 
We moreover set $k_x=k_y=2\pi/L_x$.

In Figure~\ref{fig: potential 3d}, we display the gravitational potential, for $x=y=0$, as a function of $z$ in the top panel and the associated relative error in the bottom panel.
As previously, we subtracted a constant from all potentials to ensure their minima were equal to 1. 
For both configurations, the relative error of the VGF-HybridBC approach reached its maximum at the vertical boundaries. Upper limits on the error are about $10^{-6}$ and $10^{-4}$, for $H=1.0$ and $H=3.0$, respectively. Note that the error vanishes exactly in the symmetry plane (i.e., at $z=0$). 
In comparison, the SASHA approach is slightly less accurate (by a factor of a few in terms of the relative error) than the revisited VGF-HybridBC method.

For completeness, in Figure~\ref{fig: err_xy and err_xz} we show the colour-coded relative error in the $z=0$ plane (top), and $y=3$ plane (bottom), respectively. 
The overall error was less than $10^{-5}$ (with the maximum reached near the vacuum boundaries), including larger regions where the error was exactly zero (i.e. below machine accuracy).

\subsection{Numerical convergence}

\begin{figure}
\centering
\includegraphics[width=0.9\hsize]{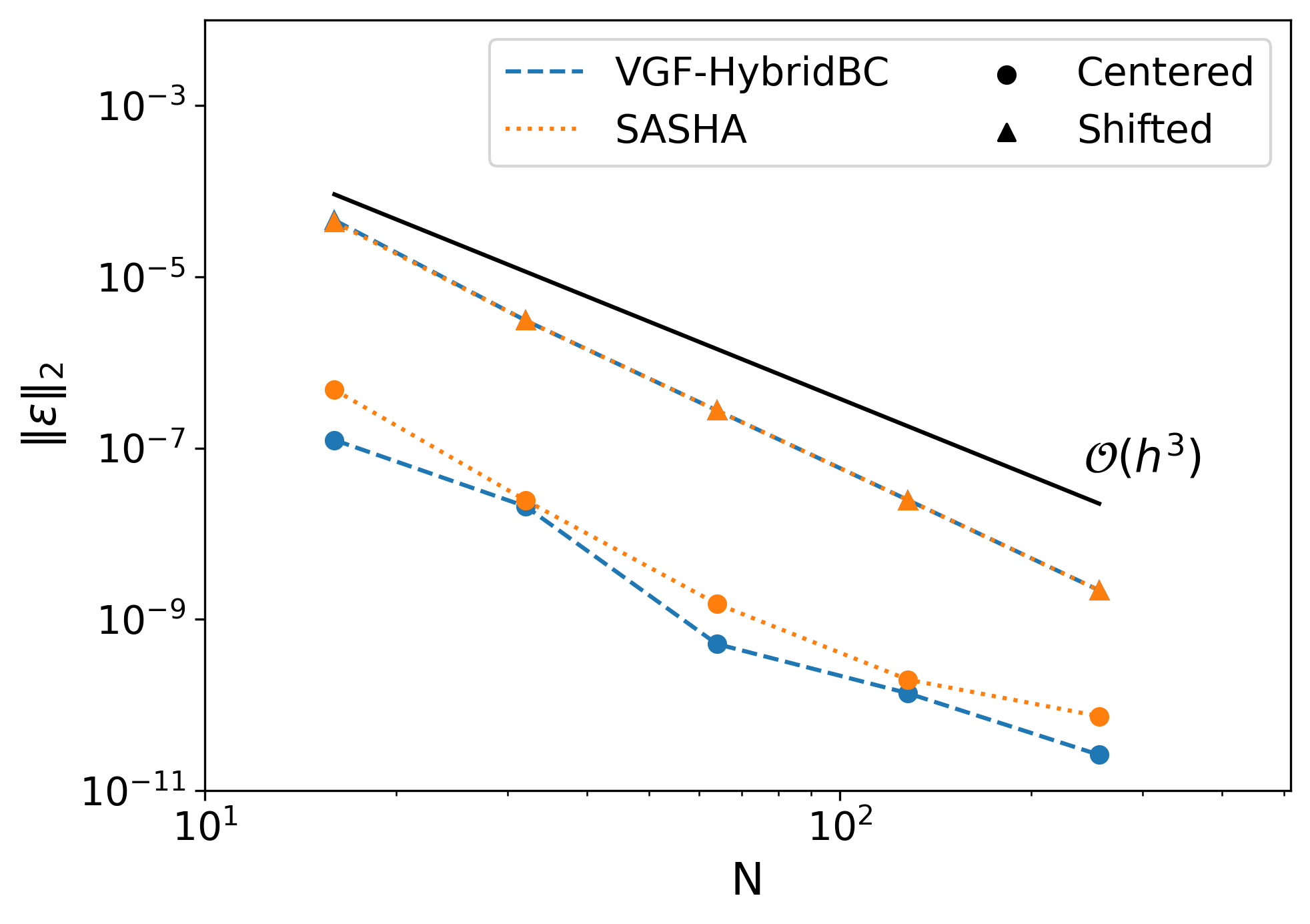}
\includegraphics[width=0.9\hsize]{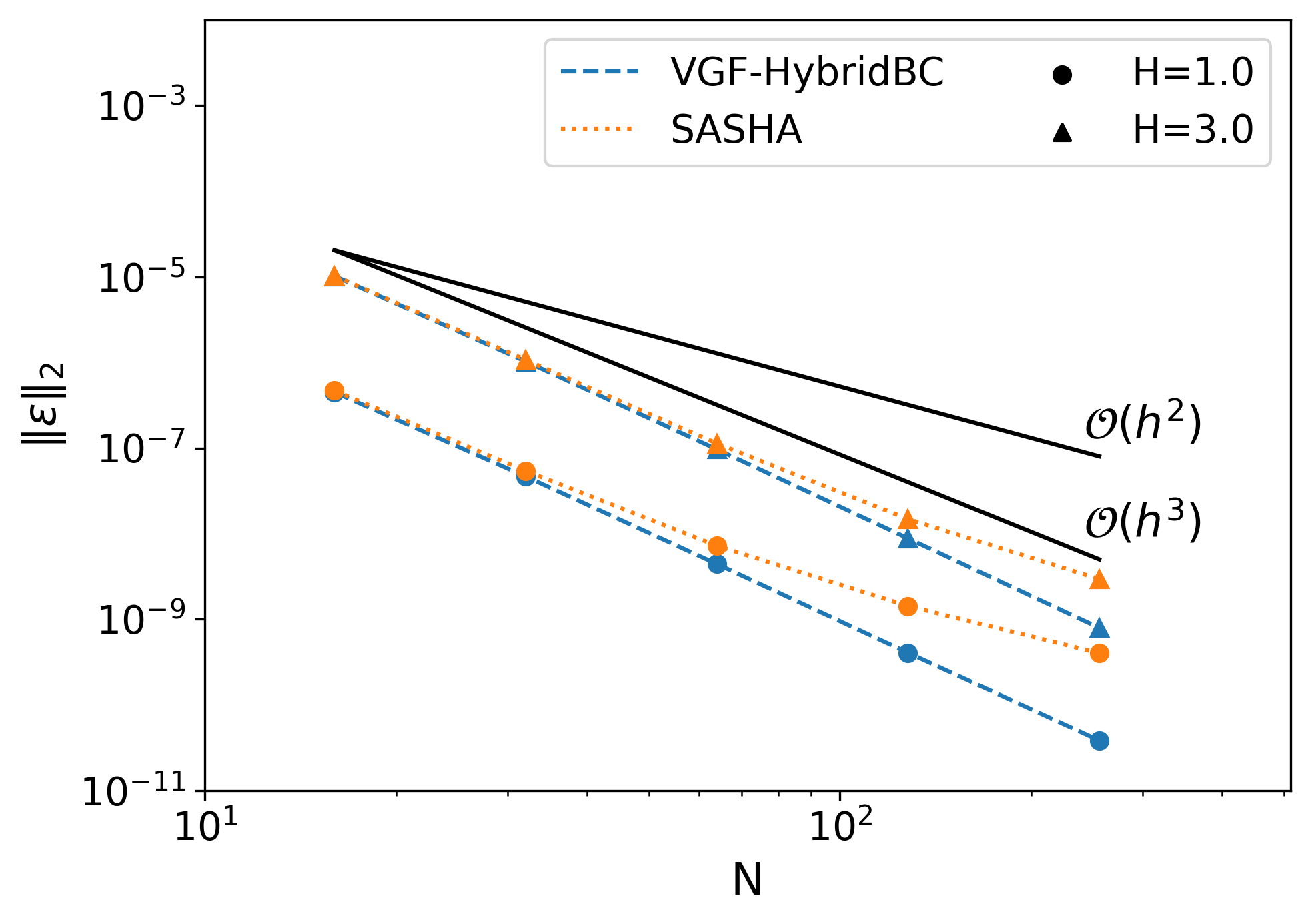}

\caption{Convergence test: 1D (top) and 3D (bottom) results.
The SASHA technique achieves second-order convergence, while the VGF-HybridBC method reaches slightly better than third-order convergence.
}
\label{fig: numerical convergence}
\end{figure}

For the numerical convergence test, we conducted 1D and 3D tests as outlined in Sections~\ref{subsec: 1D vertical test} and \ref{subsec: 3d with mixed BC}, varying the number of cells from $16^3$ to $256^3$. Figure~\ref{fig: numerical convergence} displays the $L_2$ norm of the relative error,
\begin{equation}
\left\Vert \epsilon \right\Vert_2 = \frac{1}{N} \sqrt{\sum \limits_{i,j,k} \epsilon_{i,j,k}^2 },
\end{equation}
for both 1D (top panel) and 3D (bottom panel) cases. 
We begin by analysing the outcomes of the SASHA method. 
In the 1D test, errors exhibit third-order convergence, with the highest error observed for the shifted distribution. 
Conversely, in the 3D test, this method demonstrates second-order convergence. 
Moving on to the VGF-HybridBC method, we find that in both scenarios, errors converge at a third order. 
The worst-case error is approximately $10^{-5}$, with the best-case error reaching $10^{-7}$ for $16^3$ cells. 
For $256^3$ cells, the error for all tests is consistently below $10^{-9}$ and reaches as low as $10^{-11}$ for the 3D test with $H=1$.

\subsection{Jeans instability}

\begin{figure*}
\centering
\resizebox{\hsize }{!}{\includegraphics[width=\hsize]{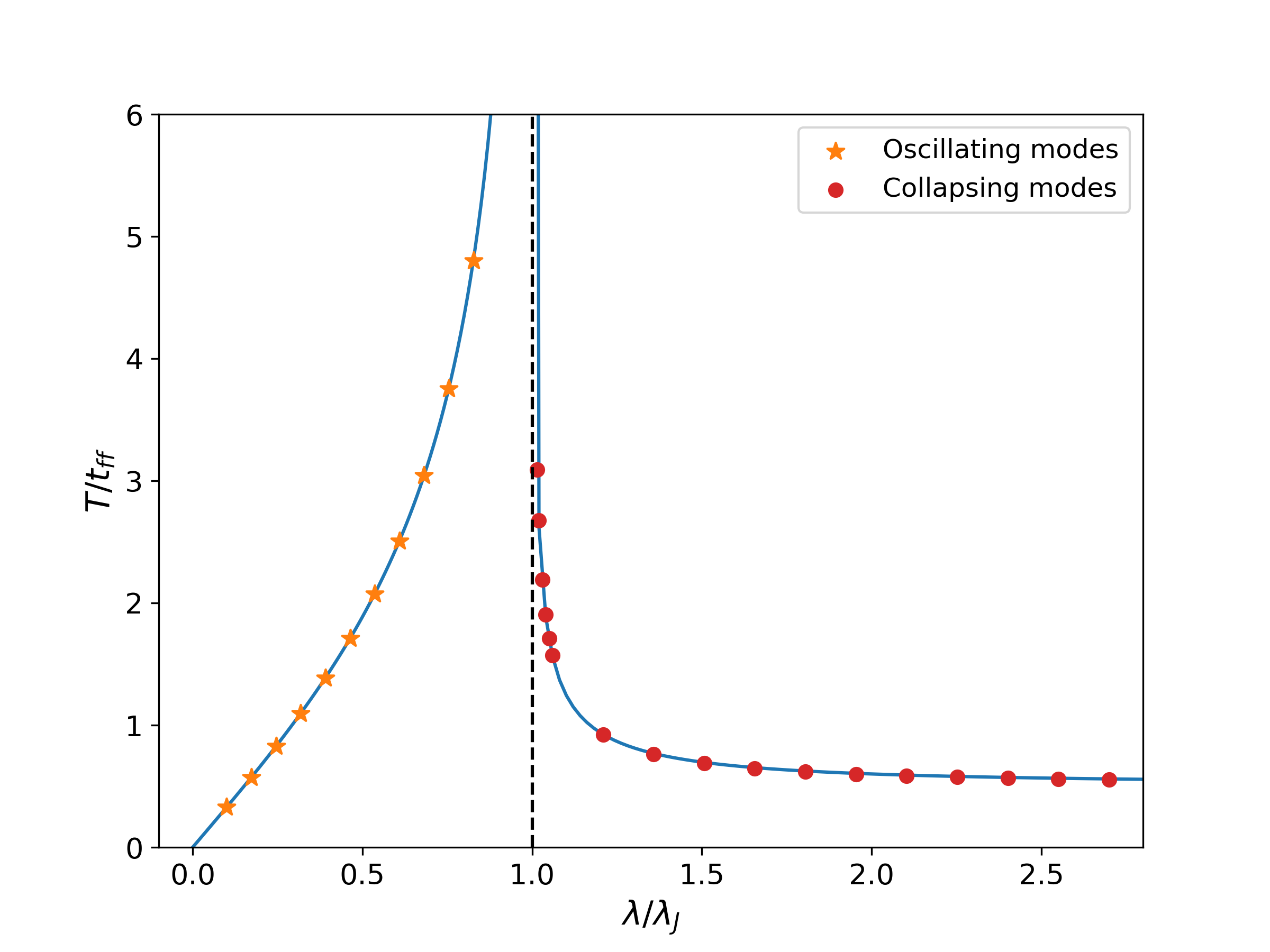}
\includegraphics[width=\hsize]{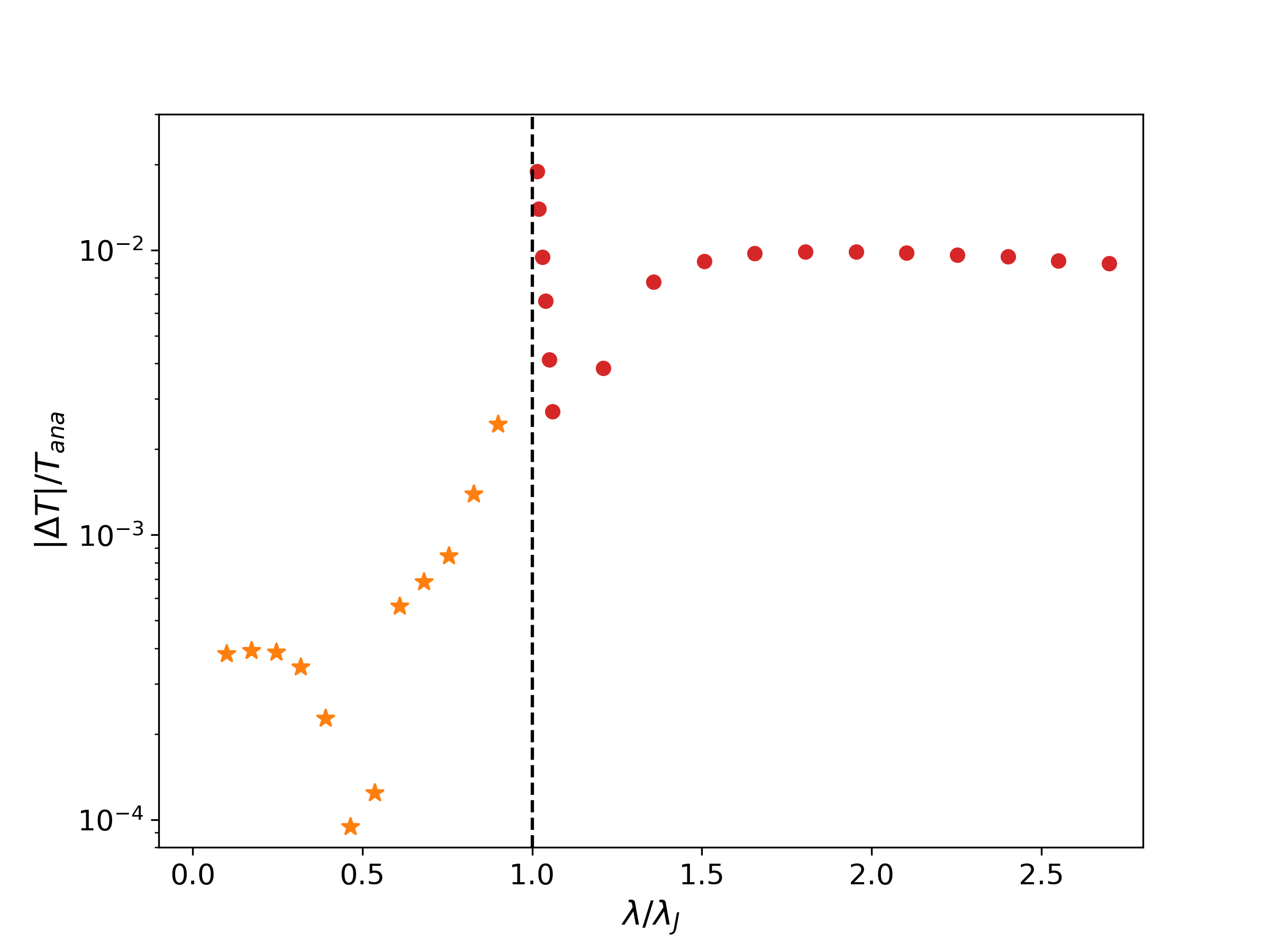} }

\caption{Left: Characteristic times of oscillatory and collapse modes with respect to the wavelength. Right: Relative error. }
\label{fig: jeans instability}
\end{figure*}

This 1D dynamical tests consists on the trigger of a gravitational instability of an uniform density distribution at equilibrium and is particularly tailored for solving the Poisson equation with periodic BC.
Therefore, we consider a 1D box with 128 cells is the $x$ direction and consider an uniform density distribution, augmented by a small sinusoidal perturbation:
\begin{equation}
\rho(x) = \rho_0 \left( 1 + A \cos{\left( \frac{2 \pi x}{\lambda} \right)} \right)
\end{equation}
with $A=0.01$.
Above perturbation can either lead to oscillating short-wavelength perturbations or to collapsing long-wavelength modes \citep{2006_hubber, 2023_mandal}, whose period and time respectively are:
\begin{equation}
\displaystyle T_{\text{osc}} = \left(\frac{\pi}{G \rho_0}\right)^{1/2} \frac{\lambda}{\left(\lambda_J^2-\lambda^2 \right)^{1/2}}  \, ; \, T_{\text{coll}} = \left(\frac{1}{4 \pi G \rho_0}\right)^{1/2} \frac{\lambda}{\left(\lambda^2-\lambda_J^2 \right)^{1/2}} \\
\end{equation}
where $\lambda_J=\sqrt{\frac{\pi c_s^2}{G \rho_0}}$ is the Jeans length.
Specifically, the collapse time is the time it takes the peak density to grow from $A \rho_0$ to $A \rho_0 \cosh{(1)}$.
As \citet{2023_mandal}, we fix $\lambda=2$ and vary the ratio $\lambda/\lambda_J$, which is equivalent to changing the sound speed.
This choice permits to have one perturbation normal mode at the initial state for all setups.

In Figure~\ref{fig: jeans instability}, we depict the normalised time (i.e., with respect to the free fall time) with respect to the normalised wavelength.
The star and circles correspond to the fitted numerical points in the oscillatory and collapsing regime, respectively, while the solid lines are the theoretical estimates.
We observe that the error between the theoretical prediction and the numerical estimate is smaller than 0.3 \% in the oscillatory regime, and about 2\% in the collapsing regime -- which validates this test.

\section{Performance estimates}\label{sec: performance}

\begin{figure}
\centering
\includegraphics[width=\hsize]{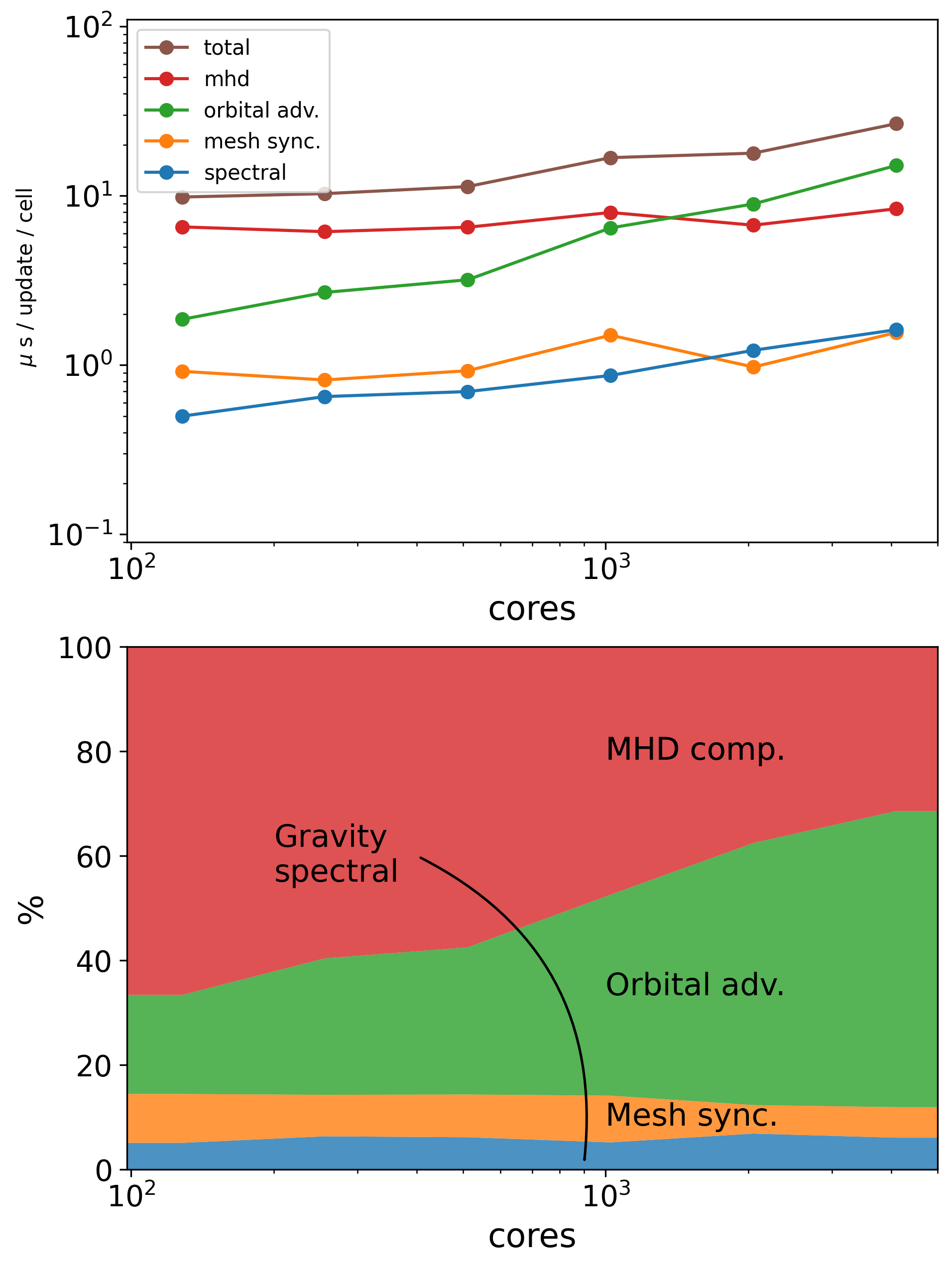}

\caption{Weak scaling study for a workload of $32^3$ cells per MPI task and using a pencil decomposition (\PDFFT).
The time spent in the spectral solver is less than 6\% of the whole runtime.
}
\label{fig: weak scaling}
\end{figure}

We have benchmarked the real-life performance of our algorithm on the \textsc{Romeo} cluster at the HPC Center of TU Dresden.
Specifically, that machine is a general-purpose NEC cluster featuring 188 nodes, each equipped with dual AMD EPYC 7702 processors (64 cores @ 2.0 GHz, multithreading enabled), 512 GB of DDR4-3200 RAM, and 200 GB of local SSD storage.
We employed the \texttt{exclusive} option of \textsc{slurm} to mitigate interferences with other jobs.

For the presented weak scaling test, each MPI task handled a fixed workload of $32^3$ cells. 
We emphasize that, for a fixed number of cores, \PDFFT performs optimally for a processor grid configuration satisfying $P_y \geq P_x$, where $P_x$ and $P_y$ denote the number of processors in the $x$ and $y$ directions, respectively. 
Additionally, $P_y$ should be set to the maximum number of logical cores per node, which is 128 for the \textsc{Romeo} cluster.
In contrast, the MHD solver achieves optimal performance with a grid decomposition that minimizes the surface-to-volume ratio.
For our setup, this implies that $P_x$ should be as close as possible to $P_y$.
To reconcile these conflicting requirements, we tested multiple processor grid decompositions for a fixed number of cores, adopting the configuration with the minimal run-time as the baseline for each core count.

In Figure~\ref{fig: weak scaling}, we present the scaling of the VGF-HybridBC method, implemented in the finite volume code \NIRVANA, from 128 to 4096 processors. 
For completeness, we compared the timings of our spectral solver with those of the orbital advection, mesh synchronization, and MHD solver.
We found that the time spent ~a) in the interpolation to the nearest periodic point and ~b) the spectral solver increases with the total number of cell points, but its share in the total computation time remains almost constant and below 6\%.

We emphasize that the timings of our solver are problem-independent: spectral methods do not require iterations that depend on the problem, and the number of operations is fixed for a given grid size, regardless of the density distribution.
Therefore we expect to maintain the obtained performance also in any kind of production runs.

\section{Conclusion}

We have presented two novel spectral methods for solving the Poisson equation, tailored for Cartesian shearing box simulations with vertical vacuum boundary conditions. 
Our first method, SASHA (Superposition Analytical-Spectral Hybrid Approach), combines an analytical solution with a spectral solution via the superposition principle. 
Our second approach, VGF-HybridBC (Vico-Greengard-Ferrando with Hybrid Boundary Conditions), leverages the free-space nature of the Green's function, yielding an analytical and regularized expression in Fourier space (Eq. \ref{Eq: Green's function regularized in Fourier}). 
This last approach allows the gravitational potential to be evaluated in a single step as a 3D convolution in Fourier space.
This enables the use of fast Fourier algorithms, achieving high performance and accuracy.

We have implemented these new methods in the Finite Volume code \NIRVANA.
To optimize performance, we integrated both methods with the \PDFFT library.
This allows for pencil decomposition, balancing the computational load between the hydrodynamic solver and the spectral solver.
Our results demonstrate excellent accuracy, with a relative error of at least $10^{-5}$ using just $16^3$ grid cells.
The methods, SASHA and VGF-HybridBC, demonstrate second and third-order convergence, respectively.
In a standard production run using orbital advection, our spectral solver accounts for less than 6\% of the total runtime, even when scaled up to 4096 processors.

\begin{acknowledgements}

This work was supported by the European Union (ERC-CoG, \textsc{Epoch-of-Taurus}, No. 101043302). 
Views and opinions expressed are however those of the author(s) only and do not necessarily reflect those of the European Union or the European Research Council. Neither the European Union nor the granting authority can be held responsible for them.
The authors gratefully acknowledge the computing time made available to them on the high-performance computer at the NHR Center of TU Dresden. 
This center is jointly supported by the Federal Ministry of Education and Research and the state governments participating in the NHR (www.nhr-verein.de/unsere-partner).

\end{acknowledgements}

\bibliographystyle{aa}
\bibliography{bibliography}

\begin{appendix}

\section{Green's function in Fourier space}\label{app: Green's function in Fourier space}

In this section, we detail the derivation of the Fourier transform of the Green's function associated with the VGF-HybridBC method:
\begin{equation}
\hat{\mathcal{G}}^L(k, k_z)  = \int\limits_{-\infty}^{\infty}  \mathcal{G}^L(z)\; e^{i k_z z} \, {\rm d}z\, .
\end{equation}
Following the definition of the Green's function in real space (Eqs. \ref{Eq: green's function original} and \ref{Eq: green's function with rect}), two cases must be distinguished:

\begin{itemize}
    \item the case $k=0$: \\
    \begin{equation}
    \begin{array}{lll}
    \hat{\mathcal{G}}^L(k=0, k_z) & = & \displaystyle \int\limits_{-L}^{L}  \frac{|z|}{2} e^{i k_z z} \, {\rm d}z\,, \\
                              & = & \displaystyle \int\limits_{0}^{L} z \cos{\left( k_z z \right)} \, {\rm d}z\,, \\
                              & = & \displaystyle \frac{ k_z L \sin{(k_z L)} + \cos{(k_z L)} -1 }{k_z^2}
    \end{array}
    \end{equation}
    
    Expanding above expression to second order in $k_z$ and taking the limit $k_z \rightarrow 0$, we find:
    \begin{equation}
    \hat{\mathcal{G}}^L(k=0, k_z=0) = \frac{L^2}{2}
    \end{equation}
    
    \item the case $k\neq 0$: \\
    \begin{equation}
    \begin{array}{lll}
    \hat{\mathcal{G}}^L(k=0, k_z) & = & \displaystyle -\frac{1}{2k}  \int\limits_{-L}^{L}  e^{-k|z|} \; e^{i k_z z} \, {\rm d}z\,, \\
                              & = & \displaystyle -\frac{1}{k} \int\limits_{0}^{L} \cos{\left( k_z z \right)} \; e^{-k z} \, {\rm d}z\,, \\
                              & = & \displaystyle -\frac{ e^{-k L} \left( \frac{k_z}{k} \sin{(k_z L)} - \cos{(k_z L)} \right) + 1 }{k^2+k_z^2}
    \end{array}
    \end{equation}
\end{itemize}

\end{appendix}

\end{document}